RESEARCH ARTICLE

# Distance Learning and Multilingual Education: A Case Study of Challenges and Pedagogical Perspectives in the Greek Border Region

**Ariadni Mandala[1], Alexandros Gazis[2,3,\*] and Theodoros Vavouras[1,4]**

[1] *Department of Theoretical and Applied Linguistics, Aristotle University of Thessaloniki, Greece*
[2] *Department of Social Sciences, Heriot-Watt University, UK*
[3] *Department of Electrical and Computer Engineering, Democritus University of Thrace, Greece*
[4] *Department of Humanities, School of Humanities, Hellenic Open University, Greece*

**Abstract:** In increasingly multicultural and multilingual societies, foreign language learning has become essential not only for communication but also for social cohesion and professional advancement. Distance education has emerged as a flexible and accessible solution, particularly for adults seeking to enhance their linguistic and intercultural competencies. This study explores the views of foreign language teachers regarding the role of distance education in promoting multilingualism, with a specific focus on culturally diverse border regions. Conducted in the Regional Unit of Evros, Greece, the research adopts a qualitative methodology based on semi-structured interviews with five language educators working in public and private education. Findings reveal that teachers recognize the potential of digital tools such as Massive Open Online Courses (MOOCs), machine translation applications (e.g., Google Translate, DeepL), and adaptive learning platforms to support multilingual learning, particularly when used as supplementary resources. However, concerns were raised about the lack of personalized feedback, limited interactivity, and the absence of culturally contextualized content on existing platforms. Teachers emphasized the importance of digital literacy, pedagogical training, and culturally inclusive design to ensure effective implementation.

The study highlights the need for targeted support for educators in border regions and calls for more locally adapted digital resources that reflect linguistic diversity. These findings offer insights for policymakers and educational technology developers aiming to improve the quality and reach of multilingual education in remote or underserved areas.

## 1. Introduction

Modern society, shaped by rapid technological progress and the globalization of the economy, is increasingly becoming multilingual and multicultural. The evolving professional landscape requires individuals to continuously adapt and develop new skills, with foreign language proficiency emerging as a critical asset for career advancement, increased earning potential, and effective communication. However, growing professional and personal commitments often limit the available time for individuals—especially adults—to participate in traditional language learning programs. In this context, distance learning emerges as a promising solution, offering the flexibility and adaptability required to integrate language education into daily life. Specifically, distance learning refers to the use of digital platforms and tools that enable remote language instruction. Examples include Massive Open Online Courses (MOOCs), machine translation applications (e.g., Google Translate, DeepL), and adaptive learning platforms designed to personalize the learning experience. For instance, a language teacher might use video conferencing tools (e.g., Zoom) to conduct live classes or assign interactive exercises through platforms such as Duolingo or Rosetta Stone, allowing students to practice and receive feedback regardless of their physical location.

The technical merit of this article is to investigate the role of distance education in promoting multilingualism, with a specific focus on its effectiveness, tools, and pedagogical challenges in the context of adult learners. The study explores the perspectives of language educators working in public and private institutions within the Regional Unit of Evros, Greece. The main aims of the study are:
1) To examine the effectiveness of distance education tools, such as MOOCs, machine translation, and adaptive learning platforms, in helping assist and develop multilingualism.
2) To assess language teachers' perceptions of the benefits and limitations of digital tools in the multilingual learning context of education and learning.
3) To explore the cultural implications of language learning in border regions and the potential of distance learning to support both linguistic and intercultural competencies.

The paper is structured as follows: Section II provides a theoretical background through a literature review on distance education and multilingualism. Section III outlines the research framework, presenting the study's aims, research questions, sample, and methodology. Section IV includes a detailed descriptive analysis of the interview responses, followed by Section V, which discusses the results across four thematic categories. Finally, Section VI presents the study's limitations and offers recommendations for future research.

## 2. Literature Review

Referring to the international literature on distance education, research indicates that students from various academic and professional backgrounds increasingly participate in distance learning programs [1, 2]. This trend is particularly evident in cases involving specialization [3], reskilling [4, 5], or the pursuit of a second degree in a new field of study [6, 7], as well as in foreign language learning initiatives. The appeal of distance learning lies in its flexibility; learners can revisit course materials at any time to reinforce concepts [8, 9], address knowledge gaps, or refresh previously acquired information [10].

For adult learners, distance education emerges as one of the most accessible and efficient forms of learning [11, 12]. Adults often face time constraints due to professional and family obligations [13], which are recognized as significant obstacles to traditional, in-person learning [14]. In such cases, the asynchronous and self-paced nature of distance education allows them to balance their responsibilities while pursuing personal and professional development [15, 16].

However, the literature also highlights several notable challenges in the context of distance language learning, particularly for adult learners. According to [14], the most commonly reported limitations include poor internet connectivity, lack of direct communication with instructors, and limited digital literacy [17]. Many learners perceive that distance education cannot fully replicate the interactive and responsive nature of traditional face-to-face instruction [18]. Additionally, several educational platforms are reported to have non-intuitive interfaces, which further hinder engagement and progress for users who are less familiar with technology [19].

Despite these drawbacks, distance learning continues to serve as a crucial bridge between learners and educators, particularly in remote or underserved regions [20-22]. As Moore's theory of transactional distance suggests, reducing the psychological and instructional gap between learner and teacher is essential for effective distance education [23, 24]. By offering continuous access to support materials, instructor feedback, and structured learning environments, digital platforms foster sustained interest and interaction, helping learners build autonomy and engage more deeply with the content [25-28].

In parallel, the concept of multilingualism has become central to contemporary educational discourse. The study in Fishman' work [29], defines multilingualism as "a condition in which a person speaks two or more languages" [30, 31], while other researchers describe it as "the use of more than two languages (or even dialects) by a speaker or a linguistic community" [32-35]. These definitions emphasize both the individual and societal dimensions of multilingualism, which frequently arise in diverse or migratory populations where groups co-exist and interact [36, 37].

Multilingualism holds immense value in today's globalized society. At the individual level, language learning strengthens cognitive abilities such as memory, problem-solving, multitasking, and even proficiency in one's mother tongue through comparative analysis [38, 39]. It also boosts learners' self-confidence [40], cultural sensitivity [41, 42], and communication skills [43, 44]. At the societal level, it fosters intercultural awareness [45], empathy [46], and social cohesion [47]. As emphasized in the European Commission's Communication (COM(2008) 566) [48], multilingualism serves as a fundamental tool for mobility[49], creativity[50], and intercultural dialogue. It enhances employability and labor market competitiveness by promoting education that is practical, inclusive, and aligned with the needs of a global economy [51-53].

Among the most prominent developments in distance language education is the rise of MOOCs. These platforms provide broad access to high-quality, university-level content, often at no cost to learners. While MOOCs generally do not confer formal degrees, they offer certificates or credits and have seen remarkable uptake. Coursera alone has attracted millions of users within just a few years [54, 55]. Butcher et al. [56] highlight their value in democratizing access to knowledge by enabling flexible, interest-driven learning across geographical boundaries [57].

Importantly, MOOCs are not merely content repositories but function as digital ecosystems [58, 59] that support peer collaboration [60, 61], open participation [62, 63], and the development of digital literacy [64]. According to recent studies, MOOCs encourage learners to construct personalized learning paths [65, 66], choose their levels of engagement [67, 68], and critically assess their progress [54, 55, 69]. This autonomy, combined with the social and participatory nature of MOOCs, aligns well with the objectives of multilingual education and lifelong learning [70].

MOOCs also make a notable contribution to linguistic diversity by providing language instruction across multiple levels and topics [71, 72]. They allow learners to explore foreign languages [73], access native-speaker content, and adapt lessons to their own pace and interests [74, 75]. This flexibility is particularly beneficial in rural or border areas, where access to language programs may be limited [76-78].

However, despite their potential, MOOCs present several challenges. Rosewell et al. [79] and Deeva et al. [80] emphasize the urgent need for standardized quality assurance frameworks, as these courses are often accessed by heterogeneous and international audiences. A lack of consistent pedagogical structure, combined with minimal instructor interaction, can lead to confusion [81], learner isolation [82], or dropout [83]. Additionally, successful navigation of these platforms often demands a high level of digital competence and strong self-regulation skills [84, 85].

Another concern is that MOOCs tend to attract individuals who are already well-educated, thereby unintentionally excluding learners with lower educational attainment or weaker language backgrounds. The use of a foreign instructional language and the diverse cultural contexts of course content may also pose barriers to full participation and understanding [86, 87].

In sum, the literature suggests that while distance education—especially through MOOCs—offers promising pathways to support multilingual learning, it must be accompanied by targeted strategies to enhance interactivity, cultural relevance, and inclusive design. Integrating local context and fostering learners' digital and intercultural skills are essential for maximizing the educational impact of these tools [88]. However, it is worth mentioning that a critical gap remains in the current literature regarding the localized and educator-centred perspectives on distance multilingual education. Specifically, while prior research has extensively examined the general benefits and limitations of tools like MOOCs, machine translation, and adaptive platforms, there are not many researchers covering how these tools are experienced and interpreted by language teachers. This is particularly evident in the case of culturally diverse and geographically remote regions. Furthermore, the specific pedagogical and intercultural challenges faced by educators in border areas, such as the Greek region of Evros, generally remain underexplored. As such, the following sections aim to fill that gap by providing qualitative insights into the real-world application of distance learning tools in promoting multilingualism and intercultural competence within such unique educational contexts.

## 3. Theoretical Research Framework

### 3.1. Aims and objectives

The present research aims to investigate and document the opinions and preferences of foreign language teachers in public and private education within the Regional Unit of Evros regarding the contribution of distance education to promoting the learning of multiple foreign languages—namely, the advancement of multilingualism. The study focuses on the use of digital tools such as MOOCs and adaptive learning platforms. Additionally, it examines how machine translation tools (e.g., Google Translate, DeepL) can support communication and language development within these digital learning environments. Specifically, it seeks to understand the potential and limitations of these tools in developing language skills within multilingual learning environments [89, 90]. Specifically, it seeks to understand the potential and limitations of these tools in developing language skills within multilingual learning environments.

To achieve this aim, the following sub-objectives have been established:
1) To evaluate the effectiveness of MOOCs in language teaching and their contribution to enhancing multilingualism.
2) To investigate the role of machine translation tools (Google Translate, DeepL) and adaptive learning platforms in supporting language autonomy and multilingual education.
3) To explore language teachers' perceptions regarding the cultural relevance of digital language learning tools, particularly in educational contexts located in border regions.
4) To assess the extent to which distance education platforms promote equal access to multilingual learning opportunities for learners with diverse educational and technological backgrounds.

### 3.2. Research questions

The study explores the perspectives of foreign language teachers in public and private education within the Regional Unit of Evros. To gain a nuanced understanding of their experiences and perceptions regarding the role of distance education in promoting multilingualism, the following research questions were developed:
1) **How do language teachers assess the effectiveness of MOOCs** in foreign language teaching and their role in enhancing multilingualism?
2) **What are teachers' views on the use of machine translation tools** (e.g., Google Translate, DeepL) and adaptive learning platforms in supporting student autonomy and language development?
3) **How do teachers perceive the role of distance education in supporting multilingualism and cultural identity** in border or underserved regions?

### 3.3. Data samples

A total of five foreign language teachers from public and private education institutions in the Regional Unit of Evros participated in the survey. The five participating teachers specialized in English, Italian, French, and German; four were women and one was a man, none held an MSc degree, and all had 10–15 years of teaching experience in public or private schools in the Evros region.

This specific group was intentionally selected, as these educators work in a border region and are well-positioned to describe and identify the obstacles and challenges they encounter in using distance education for foreign language learning and promoting multilingualism among their students. Their insights also shed light on the potential and prospects of implementing such tools, particularly in geographically and culturally complex areas like border regions.

### 3.4. Research methodology

To investigate this issue, a qualitative study based on a sample design was conducted using semi-structured interviews [91-93]. The choice of a qualitative research method allows for an in-depth exploration of the components of the phenomenon under investigation and aims to "discover the views of the population under investigation by focusing on the perspectives through which individuals experience and interpret events" [94, 95].

Qualitative research is an interpretive, dynamic, and humanistic-oriented process that emphasizes holistic documentation and critical reflection [96, 97]. While quantitative research typically relies on numerical data and statistical tools (e.g., relevance coefficients), qualitative research centres on dialogue and argumentation as its primary tools. In educational research, this approach is particularly useful for exploring teaching and learning issues, offering deeper insights that can inform improved practices[98-101]. Qualitative research is also valued for its interpretive, dynamic, and humanistic orientation, emphasizing holistic documentation and critical reflection. Compared to quantitative methods, it contributes to a deeper understanding of educational processes and can inform the improvement of teaching practices[102-105].

In this study, semi-structured interviews were chosen as the main data collection method because they enable richer, more nuanced responses than questionnaires. Interviewees are actively engaged, allowing for direct verbal interaction and the expression of thoughts and feelings, which helps the researcher gain a deeper understanding of participants' perspectives. On average, each interview lasted approximately 35–45 minutes.

As such, thematic analysis of the interview data resulted in four main codes, each representing a key thematic category:
1. Adaptive Learning: Experiences and perceptions regarding platforms that personalize language learning.
2. MOOCs and Multilingual Teaching: Views on the use and effectiveness of Massive Open Online Courses for language acquisition.
3. Machine Translation Tools: Attitudes toward tools such as Google Translate and DeepL in language education.
4. Multilingualism and Cultural Identity in the Border Area: Perspectives on how distance education supports linguistic diversity and cultural integration in border regions.

These codes formed the basis for organizing and interpreting the qualitative data, ensuring a systematic and transparent analysis.

### 3.4.1. Research Data Collection Method

In this research, interviews were used as the primary data collection instrument. Semi-structured interviews [106, 107], were conducted with five foreign language teachers working in public and private education within the Regional Unit of Evros. The interviews took place online (via face-to-face video calls) and by telephone, following a semi-structured, ten-question protocol, [108, 109]. The responses were recorded, transcribed, and analyzed using thematic analysis, based on the methodological framework of Braun and Clarke [110], and the general rationale provided in Tsiolis [111], Wolff et al. [112], as well as Salmona et al. [113].

Thematic analysis is a flexible and systematic approach that facilitates the identification, categorization, and interpretation of recurring themes within qualitative data. In this study, it was applied to capture language teachers' experiences and attitudes toward the contribution of distance education in promoting multilingualism. The analysis process followed six distinct stages as outlined by the typology in Braun and Clarke [110]:
1) Familiarization with the data through repeated reading of the transcribed interviews.
2) Systematic coding to capture key points.
3) Development of initial themes through the grouping of codes.
4) Review of themes to ensure internal consistency and significance.
5) Naming and defining themes to support the final interpretation.
6) Drafting of the analytical narrative, incorporating illustrative quotes from the interviews [114, 115].

The analysis resulted in four main thematic categories:
1) Adaptive learning
2) MOOCs and multilingual teaching
3) Machine translation tools (Google Translate, DeepL)
4) Multilingualism and cultural identity in the borderlands

Although formal ethical approval was not required due to the nature and scope of this study, all ethical principles for conducting qualitative research were strictly followed. Specifically, participation in the study was entirely voluntary, and all participants were informed in advance about the purpose, process, and use of the data. Moreover, verbal consent was obtained before the interviews, and participants were assured of the anonymity and confidentiality of their responses at all stages of the research. Lastly, data was collected and stored securely in compliance with the general data protection guidelines. As such, our study followed established ethical protocols for informed consent, transparency, and participant protection, in line with qualitative research standards.

### 3.5 Descriptive Analysis Of Interview Questions

The interview questionnaire, which was based on Iosifidis[116] and Robson [117], was structured as follows:

1) **Topic Title:** *Distance Learning as a Tool for Promoting Multilingualism: Views of Language Teachers in the Border Region*
2) **Aim:** To explore teachers' perceptions, experiences, and attitudes toward MOOCs, machine translation tools, digital adaptive learning platforms, and the intercultural dimension of distance multilingual education.
3) **Ten Interview Questions:**
   a. What is your experience using MOOCs in language teaching, and what do you consider their main benefits or limitations?
   b. Do you use, or have you used, machine translation tools (such as Google Translate or DeepL) in your teaching, or do you recommend them to your students? What is your opinion of their educational value?
   c. What is your opinion of adaptive learning platforms? Do you believe they help personalize language learning?
   d. In your view, does distance education enhance or limit the linguistic and cultural development of students in hard-to-reach areas, such as those in the Evros Regional Unit?
   e. What are the main difficulties you encounter or observe when students learn foreign languages using online tools?
   f. Do you think that distance learning facilitates multilingualism, or does it primarily reinforce dominant languages (e.g., English)?
   g. To what extent do digital teaching platforms enhance the cultural dimension of learning a foreign language?
   h. Have you observed any learning or cultural benefits from using multilingual online tools in your classroom?
   i. What would you consider to be ideal features for an educational platform designed for language learning in geographically isolated or borderline areas?
   j. In your opinion, what are the most essential steps needed to improve distance multilingual education within the Greek educational system?

## 4. Results
### 4.1. Discussion

The investigation of the opinions and preferences of foreign language teachers working in public and private education within the Regional Unit of Evros regarding the use of distance education for multilingual and foreign language learning revealed several important findings.

#### *4.1.1. MOOCs and multilingual teaching*

In relation to this thematic category, most participating teachers acknowledged that MOOCs provide access to high-quality educational materials, particularly for widely spoken languages such as English and Spanish [118, 119]. Four out of five participants reported that MOOCs are a valuable supplementary tool for language learning, especially in resource-limited settings such as frontier-dense areas [120-122]. However, they also noted that the absence of direct communication and interactivity limits the dynamic nature of such courses. Some teachers observed that while MOOCs benefit autonomous and self-disciplined learners, they are less effective for those who require structured guidance. Furthermore, the lack of personalized feedback and cultural content was seen as a significant drawback for intercultural education. In border regions, MOOCs are viewed primarily as a supportive rather than a primary instructional method. As Professor K. P. (Professor 2) remarked: *"They are useful as a supplement, not as an exclusive form of learning".* These findings suggest that while teachers value MOOCs for their accessibility and supplementary role, they emphasize the need for greater interactivity and cultural relevance.

#### *4.1.2. Machine translation tools (Google Translate, DeepL)*

This category revealed that the use of machine translation tools is widespread among both students and teachers. All participants reported using such tools at some point. Educators acknowledged their value in facilitating the immediate understanding of unfamiliar vocabulary and phrases, particularly when paired with self-correction and reflective learning practices. Most teachers considered DeepL to be more reliable than Google Translate due to its superior handling of grammatically and syntactically complex texts.

Despite these advantages, concerns were raised about the passive use of such tools without comprehension of underlying language structures, which may reduce learners' effort in producing authentic speech. All five teachers agreed that these tools can enhance vocabulary acquisition and quick comprehension, but they are not a substitute for developing productive language skills or for embedding cultural context elements essential in multilingual, multicultural education. The importance of guiding students in the correct use of these tools was emphasized [123, 124]. Teacher A. K. (Teacher 3) stated: *"It is good to start with them, but you have to learn to judge the content."* Similarly, teacher L.E. (Teacher 4) pointed out: *"Everything dictates the need to improve the level of knowledge already acquired to acquire new skills and to enable the individual to develop. Students, and especially adults, are constantly seeking to upgrade and redefine their knowledge and skills, but they are faced with the challenge of balancing personal and professional lives."*

### *4.1.3. Adaptive learning*

Opinions regarding adaptive learning platforms showed interesting variation. Only two of the five teachers had used such platforms. These two participants described the tools positively, highlighting the ability to personalize learning experiences and adapt to individual learning paces. Personalized feedback and progress tracking were considered key advantages. However, concerns were also raised about the effectiveness of these tools without proper pedagogical guidance, the lack of sufficient teacher training, and the high cost of many commercial platforms [125]. Teacher D. K. (Teacher 1) commented: *"We need well-trained teachers and the right platform. Otherwise, it doesn't work"*. The responses indicate that adaptive platforms are appreciated for personalizing learning, yet their effectiveness depends on proper teacher training and access to suitable resources.

### *4.1.4. Multilingualism and cultural identity in the border area*

Teachers unanimously agreed that in culturally diverse areas such as Soufli, Metaxades, Didymoteicho, Orestiada, Dikaia, Lavaras, and Tychero—where different linguistic and cultural identities coexist—distance education could play a crucial role in promoting multilingualism [126-129]. At the same time, they expressed concerns about the detachment of language learning from the cultural context when standardized digital platforms are used. They advocated for the development of locally adapted tools that incorporate elements of local culture. Three out of five teachers stressed that distance learning can broaden access to language instruction, but is insufficient for fostering cultural identity unless enriched with culturally relevant content. The need to strengthen both teacher and student intercultural competencies was also emphasized. As Teacher D. P. (Teacher 5) noted: *"Language is a carrier of culture. If the lesson does not convey culture, it is half a lesson."*

In summary, the analysis of the interviews indicates that teachers view distance learning as a potentially powerful tool for enhancing multilingualism, though with clear reservations. MOOCs are recognized as useful supplementary tools, while machine translation platforms increase access to language but fall short in pedagogical depth. Adaptive learning technologies show promise but require appropriate teacher training and thoughtful integration. Importantly, in border regions, combining language instruction with cultural elements is vital for fostering multilingualism that meaningfully supports learners' identity and engagement with the global educational community. As such, teachers stressed that language education must be closely linked to cultural context, especially in border regions.

The tables below (Table 1 and Table 2) present the conclusions drawn from the thematic analysis of the responses to the interview questions under the research theme *"Distance Education as a Tool for Promoting Multilingualism: Challenges and Perspectives."*

**Table 1 Methodology of interviewer responses**

| Thematic Category | Summary Conclusions | Educational Observations / Examples |
|---|---|---|
| MOOCs and multilingual teaching | MOOCs offer access to quality learning and enhance autonomy, but lack interaction and cultural dimension. | Useful as supportive tools, but not as a standalone form of teaching. |
| Automatic translation tools (Google Translate, DeepL) | Widely used. They help with quick understanding, but do not substitute critical thinking or linguistic structure. | DeepL is considered more accurate. Guidance for students in their use is necessary. |
| Adaptive learning | Shows potential for personalization and progress monitoring. However, it requires appropriate training and proper integration. | Two out of five participants have experience using it and recognize its benefits. |
| Multilingualism and cultural identity in border regions | Distance education can facilitate language access, but it does not replace the need for cultural cultivation. | There is a need for locally adapted educational content and the promotion of intercultural understanding. |

**Table 2** Grouping of interview findings

| Thematic Category | Summary Conclusions | Educational Observations / Examples |
|---|---|---|
| Linguistic diversity and use of the mother tongue | The mother tongue contributes significantly to identity and learning, but is not always supported. | Important for immigrant and refugee children. |
| Language skills in a digital environment | Digital tools contribute to the development of multilingual skills and reinforce autonomous learning. | Applications that support vocabulary and comprehension. |
| Language learning through collaboration | Collaboration enhances communication and cultural understanding. | Intercultural projects and collaborative digital platforms. |
| Language education and inclusion | Inclusion requires language support and differentiation. | Importance of supporting vulnerable groups and adapting teaching. |

## 4.2. Limitations

The choice of the sample, the data collection tool and the type of research can be a limitation, affecting the data collected and, of course, the conclusions drawn. In carrying out this research, therefore, certain issues arose and concerns were raised that need to be highlighted.

Firstly, the fact that the sample of participants used was limited and specific since it concerned only foreign language teachers serving in public and private education in the region of the Regional Unit of Evros. Then, the sample of participants used was limited and specific since it involved only five language teachers in public and private education. Certainly, the participation in the survey of a larger number of language teachers from other Primary and Secondary Education Directorates would have provided better and more faithful results to study and would have led to more accurate conclusions.

Despite the above limitations, the present research paper came to some interesting conclusions, which could form the basis for future quantitative or qualitative research.

## 5. Conclusion

Overall, this study highlights the complex and evolving role of distance education in supporting multilingualism, especially in border regions where access to traditional learning structures may be limited. Digital tools—such as MOOCs, machine translation platforms, and adaptive learning systems—offer valuable opportunities. However, their success depends largely on thoughtful implementation, including proper teacher training, curriculum integration, and sensitivity to learners' cultural backgrounds.

The findings, based on the experiences of educators in the Evros region, emphasize the need for inclusive, culturally aware, and student-focused learning environments. As educational technology continues to advance, future research and policy must address the gap between access to digital tools and their actual impact on language learning. Closing this gap is essential not only for promoting linguistic diversity but also for building intercultural understanding in today's global and multilingual world.

Several key areas emerged that merit further study. While this research gave useful insight into the views of foreign language teachers in the Regional Unit of Evros, the small sample size limits the generalizability of its findings. A broader follow-up study involving more participants from both border and non-border educational regions would improve the reliability of results and offer a deeper understanding of the topic.

While this study provides valuable insights into the role of distance learning in promoting multilingualism within a Greek border region, caution is warranted when extending these findings to other contexts. The research was based on a small group of teachers from a specific geographic and cultural area, which may limit the direct applicability of results to non-border or urban educational settings. Nevertheless, many of the challenges identified—such as the need for targeted teacher training, cultural contextualization of digital tools, and the importance of platform usability—are widely relevant. These lessons can inform the adaptation and improvement of digital language education in more centralized or urban environments.

Moreover, a promising area for future investigation is the use of artificial intelligence (AI) in distance education. Tools like chatbots, intelligent tutoring systems, and adaptive feedback are becoming more common in digital classrooms. Still, it remains unclear how effectively they support language learning. Likewise, the use of pattern recognition software to extract data and assess learner interest deserves more attention. Further research is needed to evaluate these technologies and their potential to enhance learner engagement, personalization, and autonomy in multilingual settings. Lastly, another important direction would be to examine how educators perceive and integrate AI tools into their teaching practices, especially in linguistically diverse and resource-limited environments.